\documentclass[prl,12pt]{revtex4}

\usepackage[english]{babel}
\usepackage[intlimits]{amsmath}
\usepackage[dvips]{graphicx}

\begin{document}

\title{Dynamic Phase Transitions in Coupled  Motor Proteins}
\author{Evgeny B. Stukalin and Anatoly B. Kolomeisky}

\affiliation{Department of Chemistry, Rice University, Houston, TX 77005 USA}

 \begin{abstract}

The effect of interactions on dynamics of coupled motor proteins  is investigated theoretically. A simple stochastic discrete model, that allows to calculate explicitly the dynamic properties of the system, is developed. It is shown that there are two dynamic regimes, depending on the interaction between the particles. For strong interactions the motor proteins move as one tight cluster, while for weak interactions there is no correlation in the  motion of the proteins,  and the particle separation increases steadily with time. The boundary between two dynamic phases is specified by a critical interaction that has a non-zero value only for the coupling of the asymmetric motor proteins, and it depends on the temperature and the transitions rates. At the critical interaction there is a change in a slope for the mean  velocities and a discontinuity in the dispersions of the motor proteins as a function of the interaction energy.

\end{abstract}

\maketitle

Motor proteins are active enzyme molecules that are important for molecular transport, force generation and transfer of genetic information in biological systems \cite{lodish_book,howard_book, bray_book}. They move along the rigid linear tracks by utilizing the energy of hydrolysis of ATP or related compounds, and the chemical energy is transferred into the mechanical work with a high efficiency. However, the mechanisms of the mechanochemical coupling in the motor proteins are not fully understood \cite{howard_book}.

Structural and biochemical studies of the motor proteins reveal that they consist of many domains and subunits \cite{howard_book,bray_book,kozielski97,singleton04}, and frequently these subunits also have  enzymatic activity. An example is the helicase motor protein RecBCD \cite{vonHippel02} that corrects the DNA breaks and defects by unwinding the double-stranded DNA molecules into separate chains \cite{bianco01,dohoney01,taylor03,dillingham03}. It has three protein subunits, of which two domains, RecB and RecD, also exist as independent motor proteins \cite{taylor03,dillingham03}. Experiments  indicate that  the complex motor protein RecBCD moves {\it significantly} faster than the individual RecB and RecD subunits \cite{taylor03}. For many other motor proteins the coordination between internal domains have a strong effect on the dynamic properties \cite{asbury03,zhang04}. In addition, many motor proteins work in large groups \cite{howard_book,bray_book}, although the mechanism of such coordinated motion is largely unknown. In recent {\it in vivo} experiments \cite{kural05} the transport of organelles by kinesin and dynein motor proteins have been investigated. Although the kinesins and dyneins move in opposite directions on the microtubules, it was found that they do not work against each other. Apparently, the motor proteins moving in different directions coordinate the overall transport of the organelles. These experimental findings suggest that the inter-domain coupling in the motor proteins and the interaction  between different motor proteins have a strong effect on functioning of these biological molecules. However, theoretical investigations of these phenomena are still  limited \cite{howard_book,betterton03,stukalin05}. Recently, we proposed a theoretical approach to explain the internal  interactions in the motor proteins \cite{stukalin05}, and it was successfully applied to understand the dynamics of single RecBCD helicases. The purpose of this work is to investigate the general effect of  interactions inside the motor proteins and between the molecules on the dynamic properties of the system. 

We assume that  there are two interacting  particles that move along the parallel linear tracks, as shown in Fig. 1. This model describes the motion of RecBCD helicases with two active subunits on different DNA strands \cite{stukalin05}, or it might correspond to the transport of two interacting motor proteins (kinesins, dyneins) on the parallel filaments (microtubules). The positions of the particles $A$ and $B$ are defined by  integers $l$ and $m$, respectively, on the corresponding lattices. It is assumed that the interaction between particles favor compact vertical configurations, while the potential energy of the non-vertical configurations is larger, $U(l,m) = U_0 + \varepsilon |l-m|$, where the parameter $\varepsilon \ge 0 $ specifies the interactions. This potential of interactions seems realistic for the motion of helicases \cite{vonHippel02}, where at each step of the leading  subunit the bond between two strands of DNA should be broken, and it leads to the linear dependence of the interaction energy on the distance between the subunits.

We introduce $P(l, m; t)$ as a probability to find the system in the configuration where  $A$ is at the position $l$ on the first track and  $B$ is at the position $m$ on another track at  time $t$. The dynamics of the system can be described by a set of transition rates that depend not only on the particle type, but also on the position of the particles. For  configurations $(l \pm k, l)$ [$ k \ge 1 $], the trailing particle  can move forward (backward) with a  rate $u_{j1}$ ($w_{j1}$), where $j=a$ or $ b$ corresponds to  the particle $A$ or $B$, respectively.  At the same time, the leading particle can jump forward (backward) with a rate $u_{j2}$ ($w_{j2}$). For the vertical configurations $(l,l)$ each particle can hop forward with the  rate $u_{j2}$ or it can move backward with the  rate $w_{j1}$: see Fig. 1. Note, that in our model the transition rates do not depend on the particles separation $k = |l-m|$, but only on the "type" of transition: where it leads to a more compact configuration ($k$ decreases) or a less compact ($k$ increases). This is because of the linear  potential of interaction, $U = U_0 + \varepsilon k$, and it leads to the energy difference between two consecutive configurations being equal to  $\varepsilon$, independent of the particle separation $k$. The transition rates are related via the detailed balance relations:
\begin{equation}\label{ratios}
\frac{u_{j1}}{w_{j1}} = \frac {u_{j}}{w_{j}}\exp(+ \varepsilon /k_{B}T), \quad \frac{u_{j2}}{w_{j2}} = \frac {u_{j}}{w_{j}}\exp(- \varepsilon /k_{B}T),
\end{equation}
with $j=a$ or $b$, and where $u_{j}$ and $w_{j}$ are the hopping rates in the case of no interaction between the particles ($\varepsilon = 0$).

The dynamics of the system is governed by a set of Master equations for the probability distribution function $P(l,m;t)$,
\begin{eqnarray}\label{me1}
\frac{d P(l,l;t)}{{dt}} & = & u_{a1} P(l-1,l;t) + w_{a2} P(l+1,l;t) +  u_{b1} P(l,l-1;t) \nonumber \\
& &  + w_{b2} P(l,l+1;t)  -  (u_{a2} + w_{a1} + u_{b2} + w_{b1}) P(l,l;t); 
\end{eqnarray}
\begin{eqnarray}\label{me2}
\frac{d P(l,l-k;t)}{{dt}} & = & u_{a2} P(l-1,l-k;t) + w_{a2} P(l+1,l-k;t) + u_{b1} P(l,l-1-k;t) \nonumber \\ 
& &  + w_{b1} P(l,l+1-k;t) - (u_{a2} + w_{a2} + u_{b1} + w_{b1}) P(l,l-k;t);
\end{eqnarray}
\begin{eqnarray}\label{me3}
\frac{d P(l-k,l;t)}{{dt}} & = & u_{a1} P(l-1-k,l;t) + w_{a1} P(l+1-k,l;t) + u_{b2} P(l-k,l-1;t) \nonumber \\
& & + w_{b2} P(l-k,l+1;t)  - (u_{a1} + w_{a1} + u_{b2} + w_{b2}) P(l-k,l;t).
\end{eqnarray}
At all times these probabilities satisfy the normalization condition, $\sum\limits_{l = - \infty }^{ + \infty }\sum\limits_{m = - \infty }^{ + \infty } P(l,m;t) = 1$. The solutions of the Master equations  can be found be summing over all integers $l$ and $m$ at the fixed particle separation $k$. Defining new functions, 
\begin{equation}\label{bdef}
 P_{0, 0}(t) = \sum\limits_{l = - \infty}^{ + \infty } {P(l,l;t)},  \quad P_{0, k}(t) = \sum\limits_{l = - \infty}^{ + \infty } {P(l,l-k;t)}, \quad P_{1, k}(t) = \sum\limits_{l = - \infty}^{ + \infty } {P(l-k,l;t)},
\end{equation} 
it can be shown then that in the stationary-state limit, 
\begin{equation}\label{gen_prob}
P_{0,k}=P_{0,0} (\beta_{0})^{k}, \quad P_{1,k}=P_{0,0} (\beta_{1})^{k},
\end{equation}
where
\begin{equation}\label{beta}
\beta_0 = \frac{u_{a2} + w_{b1}}{u_{b1} + w_{a2}}, \quad \beta_1 = \frac{u_{b2} + w_{a1}}{u_{a1} + w_{b2}}.
\end{equation}
These auxiliary functions   play a critical role in our analysis. When $\beta_{0} <1$ and $\beta_{1} <1$, using the conservation of probability, we obtain
\begin{equation}\label{prob}
P_{i,k} = \frac{(1 - \beta_0)(1 - \beta_1)}{1 - \beta_0 \beta_1} (\beta_i)^k, \quad i=0,1.
\end{equation}
 This means that the vertical configuration ($k=0$) is the most probable one, and the probabilities of the less compact configurations are exponentially decreasing functions of the particle separation $k$. In this dynamic phase, the particles $A$ and $B$  correlate their overall motion. From the knowledge of the stationary probabilities  and the transition rates, the dynamic properties of the system, such as the mean velocity $V$ and dispersion (effective diffusion constant) $D$ of the center of mass, can be calculated as
\begin{equation}\label{vel}
V_{CM}  = \frac{1}{1 - \beta_0 \beta_1} \left[ (u_{a2} - \beta_0 w_{a2})(1 - \beta_1) + (u_{b2} - \beta_1 w_{b2})(1 - \beta_0) \right],
\end{equation}
and
\begin{eqnarray}\label{disp}
D_{CM} =  \frac{1}{1 - \beta_0 \beta_1} \left[ \left\{ \frac{1}{2}(u_{a2} + \beta_{0} w_{a2}) - \frac{(A_{0} + w_{a2})(u_{a2} - \beta_{0} A_{0})}{u_{b1} + w_{a2}} \right \}(1 - \beta_{1}) + \right. \nonumber  \\
+ \left. \left \{ \frac{1}{2}(u_{b2} + \beta_{1} w_{b2})- \frac{(A_{1} + w_{b2})(u_{b2}  - \beta_{1} A_{1})}{u_{a1} + w_{b2}} \right \} (1 - \beta_{0})]  \right] 
\end{eqnarray}
where the coefficients $A_i$ are given by
\begin{equation}\label{Adef}
A_0 = \frac{\beta_1 (u_{a1} - u_{a2}) + \beta_0 \beta_1 w_{a2} - w_{a1}}{1 - \beta_0 \beta_1}, \quad A_1 = \frac{\beta_0 (u_{b1} - u_{b2}) + \beta_0 \beta_1 w_{b2} - w_{b1}}{1 - \beta_0 \beta_1}.
\end{equation}
The dynamic properties of the individual particles coincide with the dynamic properties of the center of mass of the motor protein cluster. In this case, it can be shown that the average distance $L$ between the particles is always finite (in units of lattice spacings),
\begin{equation}\label{loop}
L = \frac{1}{1 - \beta_0 \beta_1} \left[\frac{\beta_0 (1 - \beta_1)}{1 - \beta_0} +  \frac{\beta_1 (1 - \beta_0)}{1 - \beta_1} \right].
\end{equation}

The situation is very different when, at least, one of $\beta_{i} > 1$ ($i=0$ or $1$).  Then from Eq. (\ref{gen_prob}) it can be concluded that less compact configurations (large $k$) dominate the steady-state dynamics of the system. In this regime the particles $A$ and $B$ move independently from each other with mean velocities (assuming  $A$ is the leading particle) 
\begin{equation}\label{rates_weak}
V_A = u_{a2} - w_{a2}, \quad V_B = u_{b1} - w_{b1},
\end{equation}
and dispersions
\begin{equation}\label{disp_weak}
D_A = (u_{a2} + w_{a2})/2, \quad D_B = (u_{b1} + w_{b1})/2.
\end{equation}
The dynamic properties of the center of mass of the motor protein cluster  is given by
\begin{equation}\label{weak}
V_{CM} = \frac {1}{2}(V_A + V_B), \quad D_{CM} = \frac{1}{4} (D_A + D_B).
\end{equation}
Furthermore, the average particle-particle separation  $L$ is steadily increasing with time.  

The boundary between two dynamic regimes is determined by the condition $\beta_{0}=1$ and $\beta_{1} <1$, or $\beta_{1}=1$ and $\beta_{0} <1$, and it depends on the transition rates and energy of interaction. Using the detailed balance conditions (\ref{ratios}), it can be argued that the transition rates can be expressed as
\begin{equation}\label{rates}
 u_{j1} = u_{j} \gamma^{1 - \theta_{j1}}, \quad  w_{j1} = w_{j} \gamma^{- \theta_{j1}}, \quad  u_{j2} = u_{j} \gamma^{- \theta_{j2}}, \quad  w_{j2} = w_{j} \gamma^{1 - \theta_{j2}}, \end{equation}
where $\gamma = \exp(\varepsilon/k_{B}T)$, and $j=a$ or $b$. The coefficients $\theta_{ji}$ are energy-distribution factors that determine the effective splitting of the interaction energy between the forward and backward transitions  \cite{howard_book,betterton03,stukalin05}. In the simplest approximation, we assume that all energy-distribution factors are approximately equal to each other,  $ 0 \le \theta_{ji}\approx \theta \le 1$,  because they describe similar transitions in the motion of the individual motor proteins \cite{stukalin05}. More general situation with state-dependent energy-distribution factors can also be analyzed. Substituting Eq. (\ref{rates}) into the expressions (\ref{beta}), we obtain
\begin{equation}
\beta_0 \gamma = (\beta_1 \gamma)^{-1} = (u_a + w_b)/(u_b + w_a).
\end{equation}
Then the boundary between two dynamic phases  corresponds to the critical value of the interaction energy,
\begin{equation}\label{ecrit}
\varepsilon_c = k_B T \left| \ln \left( \frac{u_a + w_b}{u_b + w_a} \right) \right| \ge 0. 
\end{equation}
It is  important to note that the critical interaction depends on temperature, and  in the transport of the  identical particles ($A=B$) the critical interaction is always zero. This indicates that the dynamic phase transition can only be observed for the coupling of the asymmetric motor proteins.

The existence of two dynamic phases in the transport of interacting asymmetric motor proteins  can be understood using the following arguments. Consider the configuration where the particle $A$ is $k$ sites ahead of the particle $B$ and $\varepsilon=0$. The effective rate of the transition to  the configurations where two particles are separated by $k+1$ sites is equal to $u_{a}+w_{b}$, while the effective rate for $k+1 \rightarrow k$  transition is given by  $u_{b}+w_{a}$. The free energy change of making the particle configuration less compact ($k \rightarrow k+1$) can be written as  $\Delta G(0)= - k_B T \ln \left( \frac{u_a + w_b}{u_b + w_a} \right) <0$ \cite{howard_book,fisher99}, assuming  that $u_a + w_b > u_b + w_a$. If there is interaction between the particles, then the free energy change increases by the value of $\varepsilon$, $\Delta G(\varepsilon)= \Delta G(0) + \varepsilon$. The boundary between two regimes corresponds to $\Delta G(\varepsilon_{c})= 0$, and it leads to  $\varepsilon_{c}= \left| \Delta G(0) \right|$. Thus, for strong  interactions ($\varepsilon > \varepsilon_{c}$), it is thermodynamically unfavorable to make less compact configurations. The particles cannot run away from each other, and they move as one tightly-coupled cluster. For weak interactions ($\varepsilon < \varepsilon_{c}$), the favorable  free energy change of making less compact particle configuration cannot be compensated by the energy of interaction. As a result, the distance between particles grows linearly with time, and they move in the uncorrelated fashion.

The dynamic properties of interacting motor proteins  are different in two  phases, as shown in Figs. 2 and 3. The mean velocity of the center of mass changes the slope at the critical energy of interaction, while the mean velocities of the individual particles converge to a single value - see Fig. 2. The effect of the interaction is much stronger for the dispersions. As illustrated in Fig. 3, there is a jump in the mean dispersion of the center of mass at the phase boundary. In addition, the mean dispersions of the individual particles do not converge to a single value. This discontinuity in the dispersions is a clear sign of the dynamic phase transition in the system.

In order to illustrate our approach, we consider  a simplified model of the motion of the interacting motor proteins that can only step forward, i.e., $w_{a}=w_{b}=0$. This model seems reasonable for the description of RecBCD helicase transport \cite{stukalin05}, since the experiments  indicate that the backward transitions are small \cite{perkins04}. Assuming that the particle $A$ moves faster than the particle $B$ ($u_{a} > u _{b}$), the critical interaction can be written as $\varepsilon_{c}= k_{B}T \ln (u_{a}/u_{b})$. For RecBCD motor proteins, where the transition rates for subunits can be approximated as $u_{a}=300$ and $u_{b}=73$ nucleotides/s \cite{stukalin05}, the critical interaction is $\varepsilon_{c} \approx 1.4k_{B}T$. Theoretical analysis \cite{stukalin05} estimates  the energy of interaction between the subunits in RecBCD as $\approx 6k_{B}T$, implying that this motor protein moves in the strong coupling regime, in agreement with experiments \cite{bianco01,dohoney01,taylor03,dillingham03}. Using Eqs. (\ref{vel},\ref{disp},\ref{Adef}), it can be shown that for large interactions the dynamic properties of the system  are given by
\begin{equation}
V_{CM}(\varepsilon \ge \varepsilon_{c})=\frac{(u_{a}+u_{b}) \gamma^{-\theta}}{1+\gamma^{-1}}, \quad D_{CM}(\varepsilon \ge \varepsilon_{c})=V_{CM} \left( 1-\frac{2}{\gamma(1+\gamma^{-1})^{2}} \right).
\end{equation}
In the weak coupling regime, from the expressions (\ref{rates_weak},\ref{disp_weak},\ref{weak}) it can be derived that
\begin{equation}
V_{CM}(\varepsilon \le \varepsilon_{c})=(u_{a}+u_{b}\gamma) \gamma^{-\theta}, \quad D_{CM}(\varepsilon \le \varepsilon_{c})=V_{CM}/8.
\end{equation}
The jump in the dispersions at the critical interaction is equal to
\begin{equation}
\Delta D=(u_{a}/4) (u_{a}/u_{b})^{-\theta} \left[ 2 \frac{u_{a}^{2}+u_{b}^{2}}{(u_{a}+u_{b})^{2}} -1 \right] >0.
\end{equation}
Although for this simplified model the dispersion jump is always positive, it can be shown in general that the discontinuity might have any sign.

The presented theoretical analysis of the dynamics of the coupled motor proteins is based on the simplified picture that neglects many important features of the biological transport. The intermediate biochemical states, sequence dependence of the transition rates, protein flexibility have not been taken into account in this approach.  However, it is expected that these phenomena will not change the main prediction of our analysis - the existence of the dynamic phase transitions that depend on the interaction between the particles. The most crucial assumption in our approach is the assumption of the linear potential of interactions. An important question is  if the predicted dynamic phase transitions will survive for more realistic potentials of interaction between  proteins.

In summary, the effect of interaction between the motor proteins is investigated by analyzing explicitly a simple stochastic model. Using the explicit formulas for the dynamic properties, it is shown that there are two dynamic phases for asymmetric motor proteins depending on  the interaction energy. Below the critical interaction  the particles do not correlate with each other, while above the critical interaction the particles move as a tight cluster. The origin of these phenomena is the balance between the  chemical free energy change and the change in the  energy of interactions for different transitions. The critical interaction depends on the transition rates and it can be modified by changing the temperature. Our method is applied to analyze the dynamic phase of RecBCD helicases in agreement with the experiments.  This theoretical approach suggests a new way of investigating and controlling  biological transport processes at the nanoscale level.

The authors would like to acknowledge the support from the Welch Foundation (grant C-1559), the Alfred P. Sloan Foundation (grant BR-4418) and the U.S. National Science Foundation (grant  CHE-0237105).

\newpage

\noindent {\bf Figure Captions:} \\\\

\noindent Fig. 1. Schematic view of the motion of two interacting motor proteins. Transition rates $u_{ai}$ and $w_{ai}$ ($ i = 1$ or $2$) describe the motion of the particle $A$ (small circles), while $u_{bi}$ and $w_{bi}$ are the transition rates for the particle $B$ (large circles). Any configuration is specified by the integers $l$ and $m$ for the positions of the particle $A$ and $B$, correspondingly. The energy of interaction in the configuration $(l,m)$ is equal to $|l-m| \varepsilon \ge 0$. 

\vspace{5mm}

\noindent Fig. 2. Relative velocities for the coupled motor proteins as a function of the interaction energy. Solid line corresponds to the relative velocity of the center of mass of the particles, while the dotted lines are the relative velocities of the individual particles below the critical interaction.  The parameters used for calculations are: $u_a = 4$, $w_a = 0.1$, $u_b = 1$, $w_b = 0.1$ and  $\theta = 0.02 $. 

\vspace{5mm}

\noindent Fig. 3. Relative dispersions for coupled motor proteins as a function of the interaction energy. Solid line corresponds to the relative dispersion of the center of mass of the particles, while the dotted lines are the relative dispersions of the individual particles below the critical interaction.  The parameters used for calculations are the same as in Fig. 2.

\newpage

\noindent \\\\\\

\begin{figure}[ht]
\unitlength 1in
\resizebox{3.375in}{4.10in}{\includegraphics{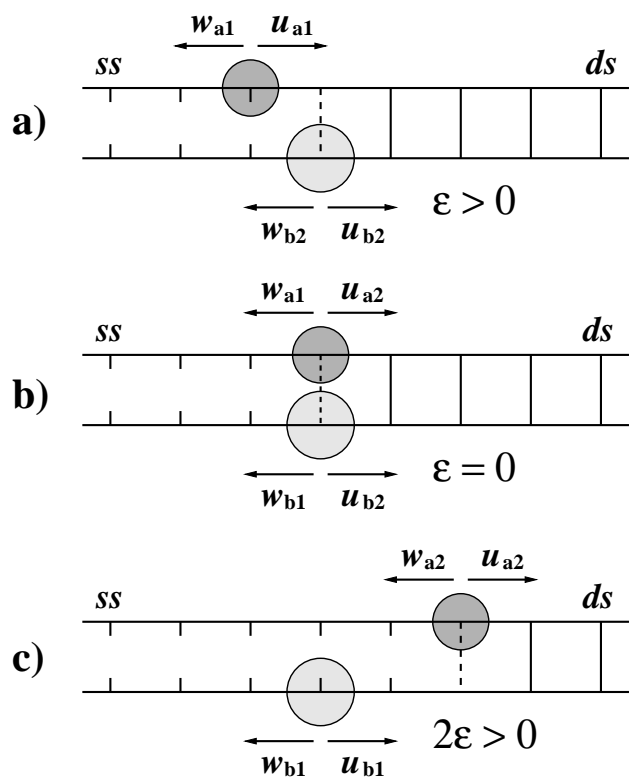}}
\vskip 0.3in
\caption{{\bf E. Stukalin, A. Kolomeisky, Physical Review Letters.}}
\end{figure}

\newpage

\noindent \\\\\\

\begin{figure}[ht]
\unitlength 1in
\resizebox{3.375in}{3.375in}{\includegraphics{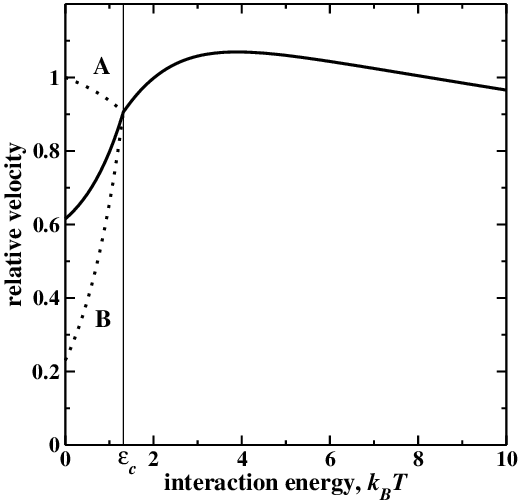}}
\vskip 0.3in
\caption{{\bf E.B. Stukalin, A.B. Kolomeisky, Physical Review Letters.}}
\end{figure}

\newpage

\noindent \\\\\\

\begin{figure}[ht]
\unitlength 1in
\resizebox{3.375in}{3.375in}{\includegraphics{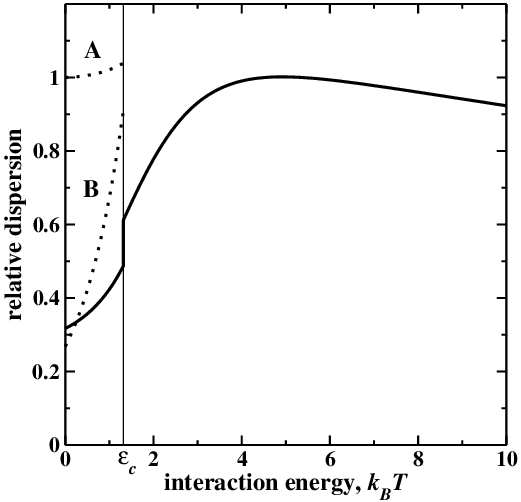}}
\vskip 0.3in
\caption{{\bf E. Stukalin, A. Kolomeisky, Physical Review Letters.}}
\end{figure}

\end{document}